# Assembly-mediated Interplay of Dipolar Interactions and Surface Spin Disorder in Colloidal Maghemite Nanoclusters

A. Kostopoulou,[a] K. Brintakis,[a,b] M. Vasilakaki,[c] K.N. Trohidou,[c] A.P. Douvalis,[d] A. Lascialfari,[e] L. Manna[f] and A. Lappas*[a]

Controlled assembly of single-crystal, colloidal maghemite nanoparticles is facilitated via a high-temperature polyol-based pathway. Structural characterization shows that size-tunable nanoclusters of 50 and 86 nm diameters (D), with high dispersibility in aqueous media, are composed of ~13 nm (d) crystallographically oriented nanoparticles. The interaction effects are examined against the increasing volume fraction, φ, of the inorganic magnetic phase that goes from individual colloidal nanoparticles (φ= 0.47) to clusters (φ= 0.72). The frozen-liquid dispersions of the latter exhibit weak ferrimagnetic behavior at 300 K. Comparative Mössbauer spectroscopic studies imply that intra-cluster interactions come into play. A new insight emerges from the clusters' temperature-dependent ac susceptibility that displays two maxima in $\chi''(T)$, with strong frequency dispersion. Scaling-law analysis, together with the observed memory effects suggest that a superspin glass state settles-in at $T_B \sim$ 160-200 K, while at lower-temperatures, surface spin-glass freezing is established at $T_f \sim$ 40-70 K. In such nanoparticle-assembled systems, with increased φ, Monte Carlo simulations corroborate the role of the inter-particle dipolar interactions and that of the constituent nanoparticles' surface spin disorder in the emerging spin-glass dynamics.

## 1 Introduction

Over the past decade, there has been a considerable progress in the synthesis of single-crystal, colloidal nanoscale magnetic particles, namely nanocrystals (NCs), because of their strong exploitation in various application fields extending from photocatalysis[1] and magnetic storage to biomedicine[2]. Complex nanoparticles (NPs) of this form are particularly appealing as the magnetic phases they carry exhibit different physical behaviour from their bulk counterparts. Enhanced or collective magnetic properties have been observed in nanoscale systems made of multiple subunits arranged in a controlled topological fashion through heteroepitaxial connections[3-5] or self-assembled in cluster-like structures. Nanoclusters with different capping agent, such as oleylamine/oleic acid[6,7], citrate[8-10] or polymers[1,11-21] have been developed. This is because their complex structure may attain collective properties[22] due to the coupling mechanisms established across the interfaced or strongly coupled material nanodomains[5,23,24]. In addition, the magnetic behavior of these complex systems may be affected by microscopic phenomena associated with the surface coordination environment, such as, canted surface spins[25], intra- and inter-particle interactions (dipolar or exchange, involving surface spins among different particles)[3,26,27] and even increased surface anisotropy[28]. Understanding of such effects is a key in the exploitation of these systems in applications strongly related to their magnetization, such as, magnetic resonance imaging (MRI) contrast enhancement[7,14,20,29], magnetic hyperthermia[30,31] and even targeted drug delivery[9,32,33].

In view of the application areas, well-known single-domain NPs, below a characteristic size (different for each material phase), exhibit unwanted, for certain technologies (e.g. magnetic data storage), superparamagnetic behavior above the so-called blocking temperature, $T_B$. While a dilute system based on such particles apparently may be easier to understand, dense systems can be a subject of debate as mutual particle interactions are not that easy to unravel. At low concentration (with respect to the dispersing medium) of such individual NPs, the inter-particle dipolar interactions are weak and the fluctuation of their magnetization is described by a characteristic relaxation time given by the Néel-Brown model:

$$\tau = \tau_0 \exp\left(\frac{KV}{k_B T}\right) \qquad (1)$$

where $\tau_0$ is the attempt time, K the effective magnetic anisotropy constant, V the particle volume and $k_B$ the Boltzmann constant.

When these nanocrystals assemble in secondary structures of high-volume fraction (of the inorganic magnetic phase with respect to the hydrodynamic volume), collective magnetic behavior is observed and the study of the magnetic dynamics is very important in order to understand the emerging properties. Generally, in these systems, the magnetic behavior is strongly dependent on factors affecting their particle magnetic anisotropy (including size, shape, crystalline phase, kind of cations involved and surface spin disorder), as well as their possible inter-particle interactions.

On the other hand, in the case of a low-volume fraction assembly of superparamagnetic NPs, the inter-particle interactions, involving superspin (i.e. single-domain particles) dipolar and surface-spin exchange interactions may be very weak. The magnetic behavior of the assembly is then governed by the intra-particle characteristics or the magnetic anisotropy of the composing particles themselves. Effectively, spin-glass behavior is the likely outcome due to the intra-particle interactions or the surface spin disorder. The latter has been observed in small nanoparticles of Ni ferrite [26], NiO[3] and maghemite[34] due to magnetic and structural disorder that arisen from broken bonds or defects on the surface of the NPs.

When the dipolar interaction strength (g) is progressively increased, the spin dynamics are dictated by an attempt time, $\tau_0$, which becomes longer[35]. Subsequently, the magnetic behavior of a nanoparticle assembly can be categorized as[36-38]: (i) superparamagnetic (weak g)[39] (ii) superspin glass (strong g)[27, 40-42], which is analogous to a canonical spin glass and (iii) superferromagnetic[43] (very strong g), in the case that the superspin moments are coupled ferromagnetically. The relaxation time then, deviates from the Arrhenius law (eq. 1) of the case (i) and follows a power-law description for the other two possibilities:

$$\tau = \tau_0 \left(\frac{T^*}{T - T^*}\right)^{zv} \qquad (2)$$

where $T^*$ is the glass transition temperature for $f \to 0$ and zv is the critical exponent, which takes values from 4 to 12 for typical spin-glass systems[44]. Furthermore, for intermediate dipolar interactions the temperature dependence of the relaxation time, $\tau$, may be approximated by the phenomenological Vogel-Fulcher law:

$$\tau = \tau_0 \exp\left(\frac{E_a/k_B}{T - T_0}\right) \qquad (3)$$

where $T_0$ represents a qualitative estimate of the inter-particle interaction energy and $E_a/k_B$ is the activation energy to overcome the barrier of the reversal of the magnetization.[45]

In the present work a high temperature polyol-based colloidal chemistry pathway, is utilized to facilitate size-controlled clustering of pure maghemite ($\gamma$-Fe$_2$O$_3$) nanocrystals. This gives rise to hydrophilic colloidal nanoclusters (CNCs). We show that the aggregation-based growth involves oriented attachment of the $\gamma$-Fe$_2$O$_3$ NPs that leads to their crystallographic alignment within the clusters. A detailed description of the interaction effects is drawn against the increasing volume fraction, $\varphi$, of the inorganic magnetic phase (i.e. from individual NPs to small and large cluster-like nanoparticle assemblies, respectively) with respect to the hydrodynamic volume. The magnetic measurements, including bulk ac/dc susceptibility of frozen aqueous dispersions and local-probe Mössbauer spectroscopy, are complemented by an elaborate theoretical approach, based on the Monte Carlo method. The influence of the inter-particle interactions, on static and dynamic properties has been explored. We show that the CNCs display weak ferrimagnetism. However, new challenges emerge from the scaling-law analysis of the frequency dispersion of the ac susceptibility and the observed memory effects, which point to a high-temperature superspin glass transition and a low-temperature surface spin-glass freezing. We decipher bear the involved microscopic interactions by simulating large assemblies of nanoparticles. Their spin-glass behaviour appears as an outcome of dipolar interactions between particles inside the nanoclusters and the parallel action of the surface spin disorder of the constituent individual NPs. We suggest that careful clarification of the magneto-structural characteristics and possible coupling effects that influence the magnetization of a colloidal assembly of nanocrystals are necessary in the engineering of functional nanoarchitectures for possible magnetically-driven application fields (e.g. MRI, magnetic hyperthermia etc).

## 2 Experimental

### 2.1 Materials

All reagents were used as received without further purification. Anhydrous iron chloride (FeCl$_3$, 98%), was purchased from Alfa Aesar. Anhydrous Sodium hydroxide (NaOH, 98%), polyacrylic acid (PAA, M$_w$= 1800), were purchased from Sigma Aldrich, while Diethylene glycol (DEG, (HOCH$_2$CH$_2$)$_2$O) of Reagent (< 0.3%) and Laboratory (< 0.5%) grades were purchased from Fisher Scientific. The absolute Ethanol was purchased from Sigma Aldrich.

### 2.2 Synthesis of Hydrophilic $\gamma$-Fe$_2$O$_3$ Nanoparticles

Colloidal syntheses were carried out under argon atmosphere in 100-mL round-bottom three-neck flasks connected via a reflux condenser to standard Schlenk line setup, equipped with immersion temperature probes and digitally-controlled heating mantles. All the reactants (FeCl$_3$, NaOH, PAA, DEG) except

Ethanol were stored and handled under argon atmosphere in a glove-box (MBRAUN, UNILab).

*a) Synthesis of iron oxide CNCs.* The γ-Fe$_2$O$_3$ CNCs were synthesized by a modified literature protocol.[46] In a typical synthesis, 0.8 mmol of FeCl$_3$ and 8 mmol of PAA were dissolved in 40 mL of DEG in a flask under anaerobic conditions maintained in the glove-box. A yellowish solution was obtained under vigorous magnetic stirring (600 RPM and a magnetic field of 250 G on the pole of the stirring bar) at room temperature. The mixture was heated to 220 °C (with ~20 °C/min) and annealed at this temperature for 1 h under argon flow. Then a 3.8 mL of NaOH in DEG hot solution (70 °C) was injected in this mixture in a single shot by using a 4 mL disposable syringe. The fast injection of the NaOH solution induced a sudden drop of the reaction mixture temperature (by 10-15 °C) and the color of the solution turned black in a few minutes. After reacting for 1 h the process stopped by removing the heating mantle and the solution cooled to room temperature. The γ-Fe$_2$O$_3$ CNCs were precipitated upon ethanol addition to the crude mixture at room temperature, separated by centrifugation at 6000 RPM for 10 min, washed three times with a mixture of de-ionized water and ethanol and finally re-dispersed in water. Further purification was accomplished by performing magnetic separations and re-dispersion in water. The second solution (stock solution), which was added at 220 °C in the starting mixture of reagents, was prepared separately from 50 mmol NaOH in 20 mL DEG and heated at 120 °C (with ~20 °C/min) for 1h. It was cooled to 70 °C and kept at this temperature till just before its injection into the starting reagents mixture. Two types of DEG grades were used, with <0.3% and <0.5% water levels to produce small and large CNCs, respectively. The same type of DEG was used for the main and the stock solutions.

*b) Synthesis of individual iron oxide NPs.* The individual NPs were synthesized by a modified literature protocol.[46] In a typical synthesis, 4 mmol of FeCl$_3$ and 4 mmol of PAA were dissolved in 36 mL of DEG (<0.5 % water) in a three-neck flask under anaerobic conditions maintained in the glove-box. A yellowish solution was obtained under vigorous magnetic stirring (600 RPM) at room temperature. The mixture was heated to 220 °C (with ~ 20 °C/min) and annealed at this temperature for 1 h under argon flow. Then 8 mL of NaOH in DEG hot solution (70 °C) was injected in this mixture in a single shot by using two 4 mL disposable syringes. The fast injection of the NaOH solution induced a sudden drop of the reaction mixture temperature (by 10-15 °C) and the color of the solution turned black in a few minutes. The γ-Fe$_2$O$_3$ NPs were precipitated upon ethanol addition to the crude mixture at room temperature, separated by centrifugation at 6000 RPM for 10 min, washed five times with a mixture of de-ionized water and ethanol and finally re-dispersed in water.

## 2.3 Characterization

*a) Transmission electron microscopy (TEM).* Low-magnification and high-resolution TEM images were recorded on a LaB$_6$ JEOL 2100 electron microscope operating at an accelerating voltage of 200 kV. For the purposes of the TEM analysis, a drop of a diluted colloidal nanoparticle aqueous solution was deposited onto a carbon-coated copper TEM grid and then the water was allowed to evaporate. Statistical analysis was carried out on several wide-field low-magnification TEM images, with the help of dedicated software (Gatan Digital Micrograph). For each sample, about 150 individual particles were counted up. All the images were recorded by the Gatan ORIUS$^{TM}$ SC 1000 CCD camera and the structural features of the nanostructures were studied by two-dimensional (2D) fast Fourier transform (FFT) analysis.

*b) X-ray Powder Diffraction (XRD).* XRD measurements were performed with a Rigaku D/MAX-2000H rotating anode diffractometer with Cu-K$_\alpha$ radiation, equipped with a secondary graphite monochromator. The XRD data at room temperature were collected over a 2θ scattering range of 5-90 °, with a step of 0.02° and a counting time of 10 s per step.

*c) Elemental Analysis.* Elemental analysis was carried out via Inductively Coupled Plasma Atomic Emission Spectroscopy (ICP-AES), using a iCAP 6500 Thermo spectrometer. Samples were dissolved in HCl/HNO$_3$ 3:1 (v/v). The concentration of the aqueous solution of small CNCs is [Fe]$_{50.2\ nm\ CNCs}$ = 44.9 ± 0.3 mM and for the large [Fe]$_{85.6\ nm\ CNCs}$ = 42.3 ± 0.3 mM).

*d) Dynamic Light Scattering and Zeta-potential Measurements.* These were carried out using a Malvern Instruments Zeta-Sizer equipped with a 4.0 mW He-Ne laser operating at 633 nm and an avalanche photodiode detector.

*e) Thermogravimetric analysis (TGA).* TGA analysis was carried out by a SDT Q600 V8.3 Build 101 TG–DTA (TA Instruments) from 20 to 600 °C with a rate of 10 °C/min in Ar atmosphere. The weight fractions of the maghemite measured were 79.8, 86.8, 92.1 % for the 12.7 nm individual NPs, 50.2 nm CNCs and 85.6 nm CNCs respectively. The volume fraction (φ) has been calculated from the weighted fraction according to the formula:

$$\frac{\rho_{\gamma-Fe_2O_3}(1-f_{m\ \gamma-Fe_2O_3})}{\rho_{PAA} f_{m\ \gamma-Fe_2O_3}} = \frac{1-\varphi_{\gamma-Fe_2O_3}}{\varphi_{\gamma-Fe_2O_3}} \quad (4)$$

where $\rho_{\gamma-Fe2O3}$, $f_{m\ \gamma-Fe2O3}$ and $\varphi_{\gamma-Fe2O3}$ are the crystal density, the weight fraction and volume fraction of γ-Fe$_2$O$_3$ nanoparticles.[47] $\rho_{PAA}$ is the average density of the organic component. The volume fractions calculated from the above equation were, φ= 0.47, 0.60 and 0.72 for the individual 12.7 nm NPs, the 50.2 and 85.6 nm CNCs, respectively.

*f) Magnetic Characterization.* The magnetic properties (dc and ac) of the samples were studied by a Superconducting Quantum Interference Device (SQUID) magnetometer (Quantum Design MPMS XL5). The measurements have been performed in purified aqueous solutions of nanoparticles and nanoclusters. The solutions were injected in a polycarbonate capsule and inserted in the magnetometer at 200 K in order to attain a frozen state. All the measurements have been performed in the temperature range where the solutions were completely frozen. The isothermal hysteresis loops, M(H), were measured at fields -1≤ H≤ +1 Tesla. The dc magnetic susceptibility as a function of temperature, χ(T), was attained down to 5 K under zero-field cooled (ZFC) and field-cooled (FC) protocols, at selected fields between 5 and 50 Oe. Before the ac measurements, a possible remnant dc magnetic field was removed using the ultralow-field option of the MPMS at 298 K. The resultant remnant field was less than 3 mOe. The frequency dispersion of the ac susceptibility (0.1 Hz≤ f≤ 1 kHz; ac field 1 Oe) was recorded as a function of temperature. Suitable phenomenological models (eqs. 1-3) were tested against the temperature evolution of the relaxation time, $\tau = (2\pi f)^{-1}$. The position of the maximum in the dissipative part of the ac susceptibility, χ''(T), was determined as the temperature at which the derivative of χ'' becomes zero. The magnetic data have been normalized to the mass of the $\gamma$-$Fe_2O_3$ as this was derived from ICP-AES analysis.

*g) $^{57}$Fe Mössbauer Spectroscopy.* The $^{57}$Fe Mössbauer spectra (MS) were collected in transmission geometry at different sample temperatures, using constant acceleration spectrometers equipped with $^{57}$Co(Rh) sources kept at room temperature. A liquid $N_2$ bath (Oxford) and a closed loop He (ARS) Mössbauer cryostats were employed for collecting spectra between 10 and 300 K. Calibration of the spectrometer was done using α-Fe at room temperature and all isomer shift (IS) values are reported relative to this standard. Fitting of the recorded spectra was done by using a recently developed least squares minimization program (IMSG09).[48] The samples were measured in powder and in solution forms; for the latter the solutions were frozen before loading them to the cryostat.

## 2.4 Monte Carlo Simulations

The Monte Carlo (MC) simulations technique, based on the Metropolis algorithm, has been used to study the macroscopic magnetic behaviour of two different models. These accounted for the spatial arrangements of the nanoarchitectures of core/surface like morphology that appeared in the experiments; namely, one which corresponds to the superparamagnetic $\gamma$-$Fe_2O_3$ NPs, well separated from one another and the second to the case where ferrimagnetic colloidal nanoclusters (CNCs) are formed. In the first model which simulates the so-called individual NPs, N identical spherical ferrimagnetic NPs of diameter d are located randomly on the nodes of a simple cubic lattice with lattice constant, a, inside a box of edge length 10 measured in units of a. The total number of NPs is N= p × (10 × 10 × 10), where p = 0.47 is the concentration of the particles in the model, equal to the volume fraction, φ, calculated from thermogravimetric analysis. In the second model, clusters of nanoparticles have been produced by dividing the lattice into eight areas with size 5× 5× 5 each and a variable particle concentration (to address the size distribution as shown by TEM) per area, but under the constraint that the total concentration is kept at the experimentally estimated φ.[49] In such a model the total concentration p= 0.60 for small clusters and p= 0.72 for large clusters is spread into eight partial concentrations, namely, 0.6, 0.7, 0.5, 0.6, 0.5, 0.7, 0.7, 0.5 and 0.72, 0.82, 0.52, 0.72, 0.82, 0.72, 0.82, 0.62, respectively. In all cases, due to the existence of the surfactant polymeric layer, it was assumed that there were no direct exchange interactions (J= 0; Scheme 1) between the nanoparticles, but instead they interacted only via dipolar forces.

We go beyond the classical model of coherent rotation of the particle's magnetization of Stoner-Wohlfarth in which each nanoparticle is described by a classical spin vector ($\vec{S}_i$ or $\vec{S}_j$). Our mesoscopic model involves a set of three classical unit spin vectors one for the core $\vec{S}_{1i}$ and two for the surface layer $\vec{S}_{2i}$, $\vec{S}_{3i}$ (Scheme 1), with magnetic moments $\vec{m}_{1i} = m_{1i} \times \vec{S}_{1i}$, $\vec{m}_{2i} = m_{2i} \times \vec{S}_{2i}$, $\vec{m}_{3i} = m_{3i} \times \vec{S}_{3i}$, respectively. In this way surface-effects were included for each i nanoparticle in the assembly. These were assumed to be coupled ferrimagnetically.

The total energy of the system is defined as follows:

$$E = -\sum_{i=1}^{N} [J_{13}(\vec{S}_{1i} \cdot \vec{S}_{2i}) + J_{12}(\vec{S}_{1i} \cdot \vec{S}_{3i}) + J_{23}(\vec{S}_{2i} \cdot \vec{S}_{3i})]$$

$$- K_C \sum_{i=1}^{N} (\vec{S}_{1i} \cdot \hat{e}_i)^2 - K_{srf} \sum_{i=1}^{N} [(\vec{S}_{2i} \cdot \hat{e}_i)^2 + (\vec{S}_{3i} \cdot \hat{e}_i)^2]$$

$$- \sum_{i=1}^{N} \sum_{n=1}^{3} H(\vec{S}_{ni} \cdot \hat{e}_h) - g \sum_{i,j=1}^{N} (\sum_{n=1}^{3} \vec{S}_{ni}) D_{ij} (\sum_{n=1}^{3} \vec{S}_{nj}) \quad (4)$$
$$i \neq j$$

The first three energy terms describe the intra-particle relations, namely the nearest-neighbor Heisenberg exchange interaction between the core spin and the two surface spins. The fourth and the fifth terms give the anisotropy energy for the core and the surface (with $\hat{e}_i$ being the anisotropy easy-axis direction), while the sixth term is the Zeeman energy (with $\hat{e}_h$ being the direction of the magnetic field). The last term describes the dipolar interactions among the spins, where $D_{ij}$ is the dipolar interaction tensor.[49] The parameters entering equation (4) are the dipolar energy strength $g = \mu_0 \, m^2/4\pi a^3$ (where m= $M_S$V is the mean particle magnetic moment, i.e. averaging the magnetic moments of the core and the shell), the intra-particle exchange energy among the core spin and the surface spins $J_{12}$, $J_{13}$, $J_{23}$, the anisotropy energy of the core $K_C$ and the surface $K_{srf}$, the external field $\mu_0$H. The thermal energy is $k_B$T (where T is the temperature).

The energy parameters in the equation (4) are based on the bulk values of maghemite ($M_S = 4.2 \times 10^5$ A/m and $K_C = 5 \times 10^3$ J/m$^3$), and their modifications are established considering the nanoparticles morphology (e.g. reduced symmetry and reduced size) using a mean field approach. Accordingly, the values of the intra-particle exchange energy among the core spin and the surface spins were taken as $J_{12}$= -7.77, $J_{13}$= -1.35, $J_{23}$= -0.091 and the anisotropy energy of the core as $K_C$= 0.1, while that of the surface as $K_{srf}$= 2.5 to 3.5, since it is expected to be more than one order of magnitude larger than that of the core. The energy parameters are normalized by the factor 10× $K_C$, so they become dimensionless.

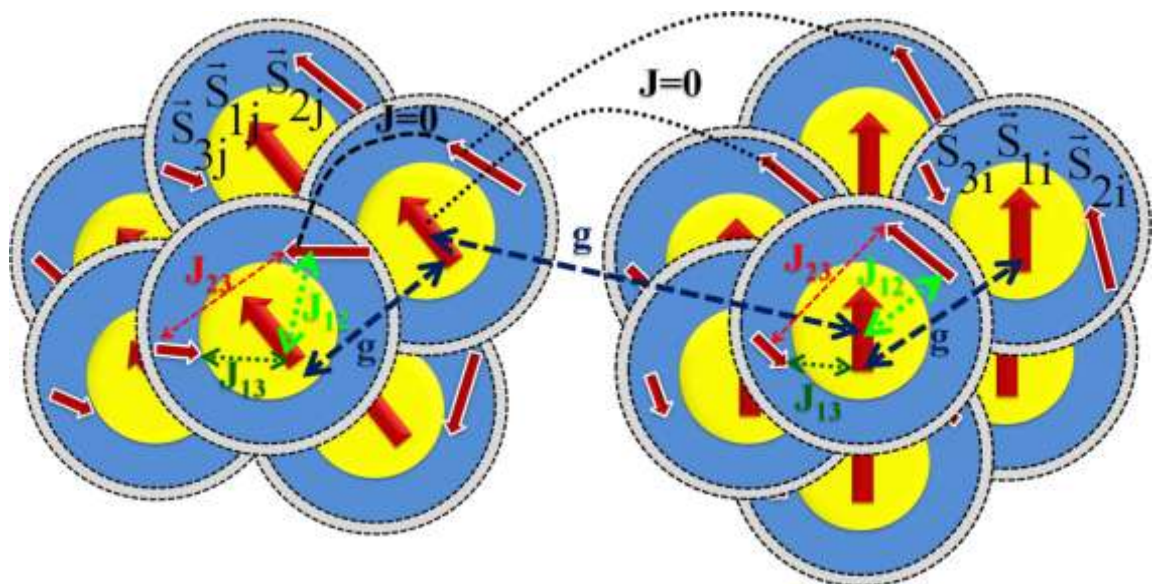

**Scheme 1.** Schematic representation of the spin structure and the intra-particle as well as inter-particle interactions in the nanoclusters. For an explanation of the labels, refer to §2.4. Outermost grey colouring: surfactant polymeric layer.

Additionally, the dipolar strength energies (g) and magnetic moments (m) have been calculated based on the relative experimental values of the particle concentration (p) and the saturation magnetization ($M_S$) of the nanoarchitectures for the three different samples. For the dipolar energy calculation the Ewald summation technique has been implemented that takes into account the long-range character of the dipolar interactions, using periodic boundaries in all directions.[50] As a result we have set g= 0.884 for p = 0.47, g = 0.955 for p = 0.60 and g = 0.865 for p = 0.72.

The simulations were performed at a given temperature and applied field, while the system was allowed to relax towards equilibrium for the first 500 Monte Carlo steps per spin. Then thermal averages were calculated over the subsequent 5000 steps. The results were averaged over 10-20 samples with various realizations of the easy-axes distribution and different spatial configurations for the nanoparticles.

## 3 Results and Discussion

### 3.1 Morphology, Crystallinity, Chemical Nature and Size-tunability

The CNCs were synthesized by a one-step high-temperature polyol-based chemical protocol.[46] Their formation relies on a two stage growth model that involves: (i) the nucleation of the iron oxide NPs in a supersaturated solution resulting from the solvent-mediated hydrolysis of $Fe^{3+}$-reagent and (ii) their controlled aggregation into larger entities. Quite monodisperse iron-oxide CNCs were formed with DEG, as a high-boiling point solvent and PAA as a capping agent. In this procedure DEG's role is also complemented by its activity as a reducing agent for the metal species.[10] The PAA was employed as a surfactant in order to control the morphology and stability of the final nanoparticle assemblies; it is adsorbed on the surface of the pre-formed NPs after their nucleation and acts as a stabilizer, regulating their size and shape evolution. The polyelectrolyte's deprotonated functional group, -COO⁻, strongly coordinates the surface of the NPs making them negatively charged. The sensitive balance of such electrostatic repulsive forces against those of magnetic origin (vide infra) determines the size and the morphology of the CNCs.

This reaction scheme is sensitive to different parameters of the hot injection process, namely, the quantity of the water in the reaction scheme, as well as the alkalinity of the reductive solution. Small differences in these variables can give rise to CNCs of different size. Utilizing exactly the same parameters, except that of the water content, which was altered only by employing two different grade solvents (§2.2.a) bearing variable water quantity (and stored under anaerobic conditions), nanoclusters with two different sizes were prepared. In addition, PAA-coordinated, individual, quite spherical NPs with size comparable to that of the nanocrystals composing the CNCs, were also synthesized. The purpose was to utilize them as a reference system that would allow comparisons to be drawn against the magnetic behavior of their assembled counterparts.

The narrow size distribution and good stability of the CNCs in aqueous suspension, excluding aggregation between them, were supported by dynamic light scattering (DLS) experiments (Fig. S1a and Table S1). Bright-field TEM images for the samples are shown in Figures 1a-c, conferring a significant degree of size/shape homogeneity. Both CNCs samples have a flower-like, quite spherical shape without aggregation between units. The individual NPs entail units of about d= 15.7 ± 3.0 nm (Fig. 1d), while each cluster is an assembly of small NPs, with no isolated particles left out of the aggregate. The average diameter, D, of the cluster entities as determined by TEM are 50.2 ± 5.4 and 85.6 ± 13.3 nm (Figs. 1e, f). It is worth noting that a similar reaction, however, carried out with all the reagents stored under non-anaerobic conditions, results in larger diameter (120.1 ± 18.7 nm) CNCs;[51] it postulates the influence of the excess moisture level as a size-enhancing growth parameter for the nanocrystal assemblies.

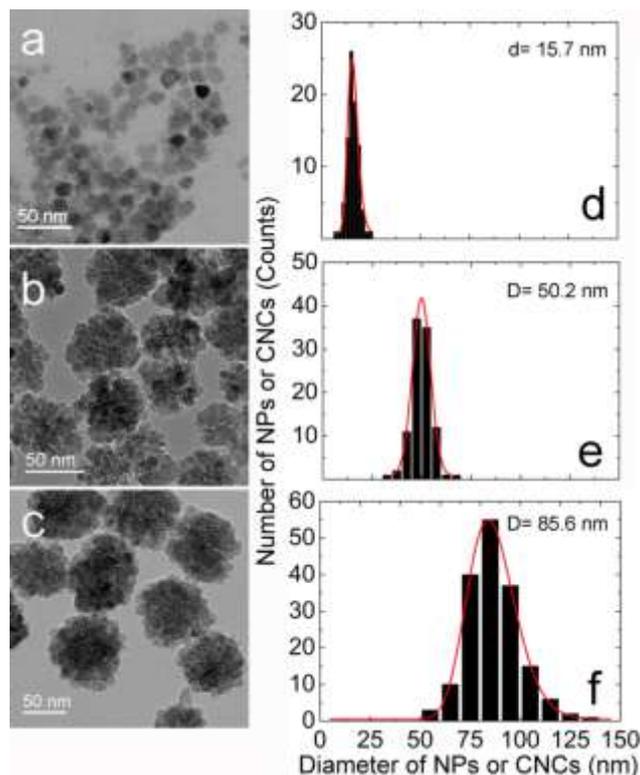

**Fig. 1** Representative low magnification bright-field TEM images of the individual NPs (a), 50.2 nm (b), 85.6 nm (c) CNCs, as well as the corresponding size distributions by TEM (d-f).

Powder X-ray diffraction (XRD) patterns of the samples (Figs. 2a-c) are attributed to a cubic spinel iron oxide structure, of either the magnetite ($Fe_3O_4$) or maghemite ($\gamma$-$Fe_2O_3$) phase. Analysis of the Bragg reflections by the Scherrer equation suggests that the colloidal nanoparticle assemblies, are constituted of a number of small NPs of d= 12.7 ± 1.0 and d= 11.6 ± 0.5 nm average diameter for the small and the large CNCs, respectively. The as-determined diameter for the individual NPs was found to be of comparable size, d= 12.7 ± 1.2 nm (Fig. 2c).

The identification of the chemical nature and especially the phase purity of the inorganic particles are important attributes when applications are sought. For this purpose we employed [57]Fe Mössbauer spectroscopy as the only reliable experimental technique to distinguish between the magnetite and the maghemite type of iron oxides. [57]Fe Mössbauer spectra (MS) recorded for the two CNCs samples at 10, 77 and 300 K show similar features at each temperature studied (Fig. S2). All the spectra were magnetically split, with increased line broadening at 300 K. Overall, these results (Table S2) constitute identifying features of ferric iron oxide nanocrystal assemblies and in particular those of the maghemite ($\gamma$-$Fe_2O_3$) stoichiometry, with their magnetic properties influenced by thermal agitation[52, 53, 55,56].

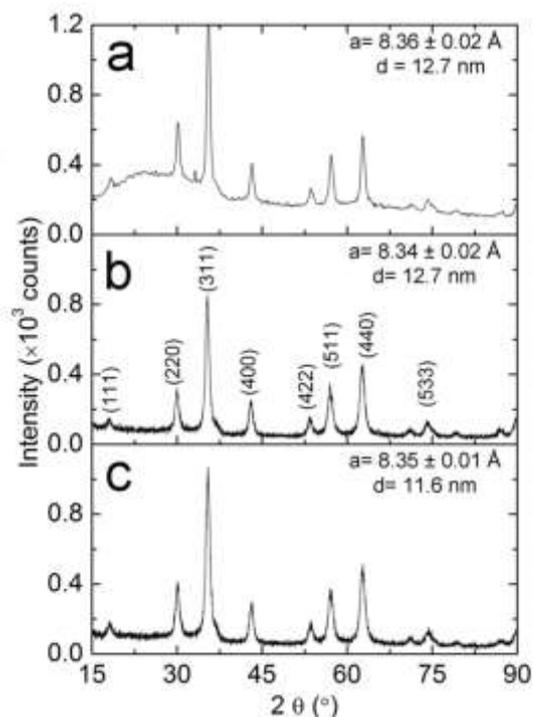

**Fig. 2** XRD powder patterns for the 12.7 nm iron oxide individual NPs (a) and the CNCs with average diameters 50.2 nm (b) and 85.6 nm (c), respectively. The small and large CNCs are composed of NPs with d= 12.7 nm and d= 11.7 nm, correspondingly. Indexing was possible on the basis of the cubic spinel structure of either bulk $\gamma$-$Fe_2O_3$ (a= 8.35 Å; ICDD 00-039-1346) or $Fe_3O_4$ (a= 8.40 Å; ICCD 00-019-0629).

Looking into the aspects of the CNCs' growth helps assessing their physical response. The purposeful choice of the surfactant renders such nanoparticle assemblies negatively charged, with good colloidal stability. Indeed, the measured z-potential was found to be -65.4 ± 10.8 and -50.0 ± 6.5 mV for the small and the large CNCs (Fig. S1b), prepared with higher and lower water content, respectively. We postulate that the relative strength of the electrostatic forces between the pre-formed, charged NPs is a major driving force that determines the CNCs' size. That is to say, stronger Coulombic repulsion amongst the NPs leads to their increased separation in the DEG-PAA liquid medium, with effect in the formation of smaller diameter CNCs. In this view, the influence of the water in the growth is justified if we consider the higher affinity of the water molecules to coordinate stronger to the surface metal cations than that of the carboxylate groups of the PAA.[54] As a result, a reduced density of polyacrylate functional groups on the surface of the as-formed NPs renders them less charged, with raised possibility to come nearer in the colloid and thus controllably assemble in larger entities.

### 3.2 Growth by Oriented Attachment

The topological arrangement of such clusters is provided by the HRTEM images and the calculated FFT patterns taken from individual CNCs (Figs. 3b-c, 3e-f). Distinct diffraction spots (Figs. 3e, f) are identified instead of diffraction rings, which are found for individual NPs (Fig. 3d) or randomly oriented polycrystalline assemblies of nanoparticles[33]. The spot pattern of the FFT resembles that of a single-crystal cubic spinel structure, but with a weak broadening due to a slight spatial misalignment between the NPs within each assembly.

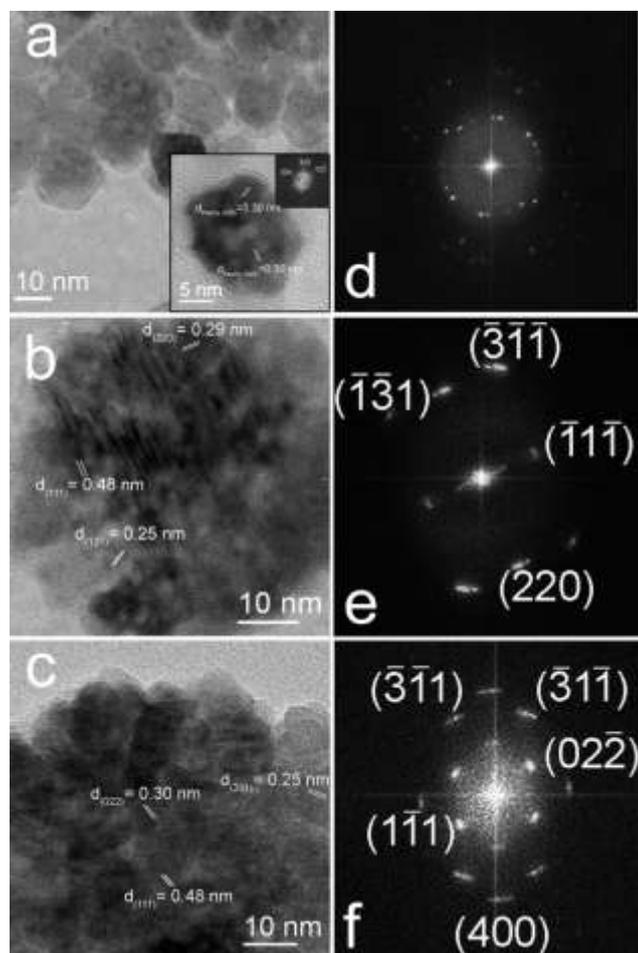

**Fig. 3** Representative HRTEM images and the corresponding FFT patterns of individual 12.7 nm NPs (a, d), 50.2 nm (b, e) and 85.6 (c, f) CNCs samples. The zone-axes are [$\bar{1}11$], [$\bar{1}12$] and [011] for the panels a, b and c, respectively. Inset in a: HRTEM image and the corresponding FFT pattern of an individual nanoparticle.

The single-crystalline like CNCs have been spontaneously formed in order to minimize their total energy. The nucleation and the growth of the NPs in a solvent such as glycol are slow enough and provide adequate time for the entities to reorient at a suitable configuration that leads to a weak misalignment amongst them during the formation of their assemblies.[13,57] Comparative studies involving such secondary nanoparticle structures formed in glycols with different reductive ability, including, ethylene (lower reducing ability) and polyethylene glycol (higher reducing ability), have demonstrated efficient oriented attachment of the comprising NPs in the case of the solvent with the lower reductive capacity.[13] Oriented attached in this form, may provide an interesting pathway for developing nanoarchitectures with collective properties and enhanced magnetic response (e.g. strong magnetic anisotropy and high magnetization) as it has been reported for 40-60 nm $CoFe_2O_4$ nanocrystal aggregates.[55]

### 3.3 Mössbauer Spectroscopy and the CNCs' Room Temperature Ferrimagnetism

The Mössbauer spectroscopy, except its contribution in the determination of the chemical nature of the NPs, gives useful information for the magnetic state of the system. The 300 K MS of the dried CNCs display magnetically split absorption lines, without any hyperfine magnetic field ($B_{hf}$) collapse, indicating ferrimagnetic-like behavior for the powder samples (Fig. 4b, c). The latter is also in agreement with the non-zero coercive field, $H_C$, derived from the SQUID magnetometry at the same temperature (Figs. 4e, f). Surprisingly, the MS of the individual NPs powder (Fig. 4a), showed significantly broader magnetically split components, with lower average $B_{hf}$ values compared to the corresponding values found for the CNCs MS (Table S2) that are in addition, superimposed on a non-magnetically split one. These results suggest the coexistence of two major microscopic mechanisms. The magnetically split component may be attributed to the ferrimagnetic-like behavior due to the intra-cluster characteristics (e.g. dipolar interaction between the particles and intra-particle exchange interactions) of the CNCs. The second mechanism, observed for the individual NPs only, entails significant diminution of the $B_{hf}$ in the MS, reminiscent to the $B_{hf}$ collapse commonly observed for small, isolated nanoparticles.[52, 56-59] The MS of the latter are dominated by thermal agitation-driven magnetization reversal that renders the nanoparticles' characteristics superparamagnetic (within the characteristic Mössbauer measuring time-scale ~$10^{-8}$ s). The different line broadening of the MS for the individual NPs can be attributed to a weaker and with a wider strength-spread inter-particle interaction, as well as to differences in the surface magnetic anisotropy and spin disorder.[53]

The appearance of the room temperature ferrimagnetism of the CNCs indicates that this magnetic material is fundamentally different from the superparamagnetic architectures of earlier studies.[8, 10, 13, 19, 46]

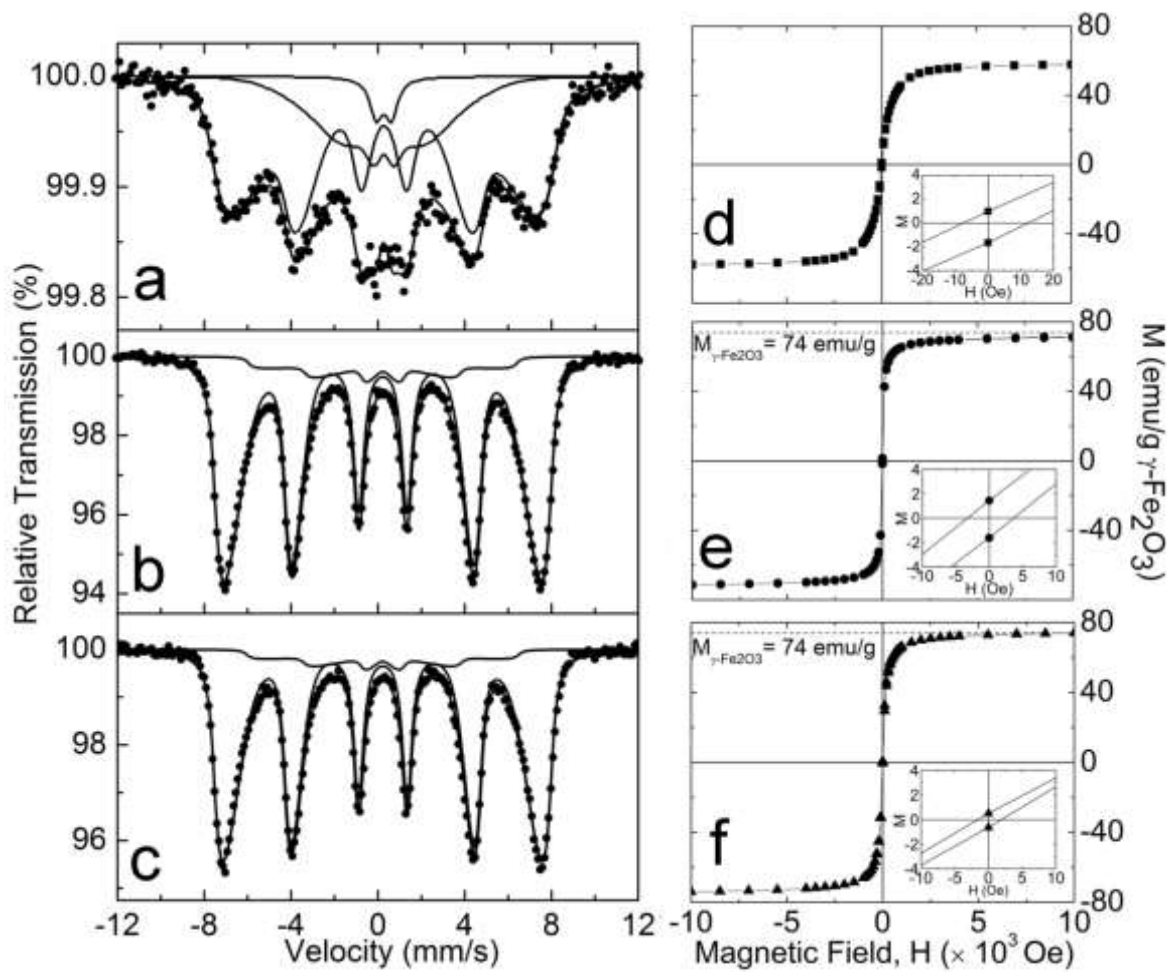

**Fig. 4** $^{57}$Fe Mössbauer spectra and magnetization curves of the dried powders at 300 K for the 12.7 nm individual NPs (a, d), the 50.2 nm CNCs (b, e), 85.6 nm (c, f) CNCs samples. Lines over the Mössbauer data are the multicomponent fits performed on the basis of underlying mechanisms discussed in the text.

### 3.4 Probing the Magnetic Interactions in the CNCs

The frequency dispersion of the ac susceptibility for aqueous frozen solutions of the individual NPs and the CNCs has been measured down to 5 K. The temperature dependence of the imaginary part of the susceptibility, $\chi''(T)$, is shown in Fig. 5 (the real part $\chi'(T)$ is plotted in Fig. S3). It is worth noting the presence of two maxima for both types (individual or clusters) of investigated colloidal nanoarchitectures. A sharper peak is recorded at low temperatures ($T_f \sim 40$ K), while a broader one at higher temperatures ($T_B > 120$ K). Similar dynamic response has also been observed in other nanoparticle-based materials, such as $\delta$-$(Fe_{0.67}Mn_{0.33})OOH$[60], nickel ferrite[26, 61], $NiO$[3], $Co_{50}Ni_{50}$[62].

Importantly the frequency dependence of the $\chi''(T)$ maximum permits the evaluation of the temperature dependence of the relaxation time $\tau$, while it provides information on the underlined spin dynamics and the origin of the two magnetic regimes. We established the validity of the available phenomenological-laws (eqs. 1-3) after successful data-fitting attempts, which give reasonable – with respect to the literature – physical parameters. Peculiarly, the frequency dependence of the $\chi''(T)$ maximum at $T_B$, which could be attributed to the blocking temperature[60], does not follow the Arrhenius law. This is because the fit values of $\tau_0$ (Table 1; Fig. S4b) were found to be uncommonly small ($\tau_0 < 10^{-19}$ s) for non-interacting superparamagnetic particles (typically, $\tau_0 \sim 10^{-13}$ s).[63] The derived large values for the activation energy ($E_a/k_B > 5000$ K), together with the short $\tau_0$, imply that the channel of inter-particle interactions must play an important role in both types of samples. In view of this, the phenomenological Vogel-Fulcher law has been utilized to fit the frequency dependence of the $\chi''(T)$ maximum. The resultant values for the $\tau_0$ ($\sim 10^{-9}$- $10^{-7}$ s) (Table 1; Fig. S4d) are comparable to those met in the literature for particles with intermediate-strength dipolar interactions.[60, 61] At the same time, a power-law, scaling analysis results in reasonable values for the $\tau_0$ ($\sim 10^{-11}$-$10^{-7}$ s) and zv (6.5-10) (Table 1; Figs. 5e, g, i), in agreement with those reported in the literature for superspin glass systems, involving strong dipolar interactions.[35, 40, 64]

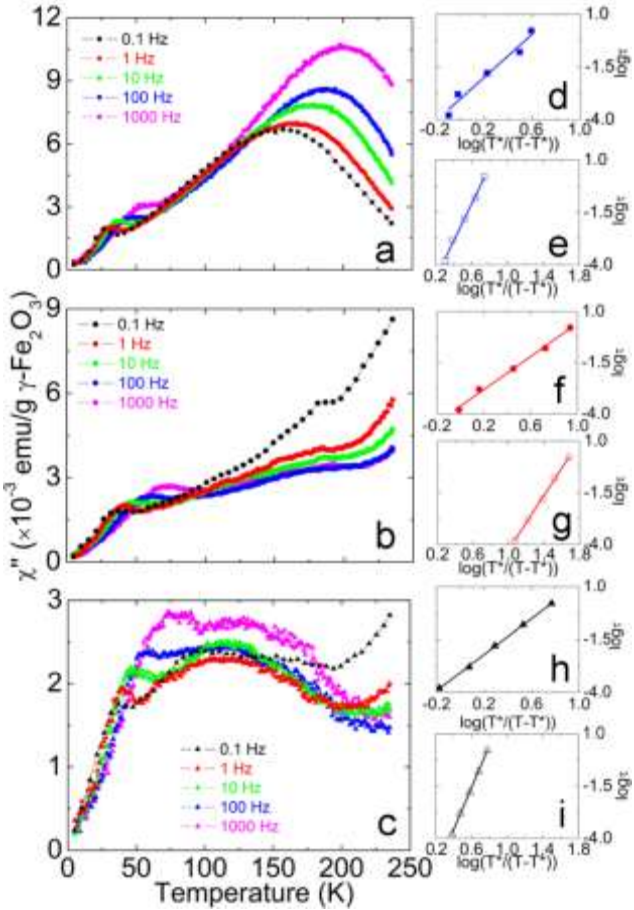

**Fig. 5** Frequency dependent imaginary part, $\chi''(T)$, of the ac susceptibility, for the 12.7 nm individual NPs (a), the 50.2 nm (b) and 85.6 nm (c) CNCs samples in solution form. Power-law, scaling analysis of the frequency dispersion of the low (filled symbols) and high (open symbols) temperature $\chi''(T)$ maxima (see text, eq. 2) (d- i).

In order to distinguish whether the Vogel-Fulcher or the power-law provide a more suitable phenomenological description of the dynamics, to a first approach one can estimate the relative variation of the $\chi''(T)$ peak-temperature position per frequency decade, $\psi$ (= $\Delta T/(T\log f)$), known as Mydosh parameter. The calculated values of the $\psi$ for the individual NPs, the small and the large CNCs, are 0.050, 0.015 and 0.047, respectively. They all fall in the range that predicts a spin-glass behaviour (0.005< $\psi$ <0.05).[65] In order check experimentally the possibility of a spin-glass freezing at $T_B$, the presence of memory effects was investigated. A reference susceptibility curve was recorded for the small CNCs under a ZFC procedure (H= 5 Oe), while the memory curve was measured as previously but after having kept the sample at 110 K for $10^4$ sec. A decrease of the susceptibility is observed in the aging curve (Fig. 6), with an onset complying with the observed plateau in the FC $\chi(T)$ (Fig. S5) suggesting a transition to a superspin glass state below the $T_B$.[66] The depth of the $\Delta\chi$ (= $\chi_{wait} - \chi_{ref}$) dip is $\sim$15$\times$ $10^{-4}$ emu/g $\gamma$-Fe$_2$O$_3$. Memory effects of this type have been observed in superspin glass systems in which strong dipolar interactions are incorporated, such as in the dense nanoparticle systems of $\delta$-(Fe$_{0.67}$Mn$_{0.33}$)OOH[60] and Fe$_3$N[66]. In accord with the latter, is the deviation from the linearity of the $T_B$ with respect to the applied dc magnetic field, H (Fig. 7a, b).[41, 60] Such evidence for a superspin glass state corroborates the power-law as a favorable model. In this case the extracted $\tau_0$s become somewhat shorter as the volume fraction, $\varphi$, increases (Table 1), a likely indication for raised inter-particle interactions.[35]

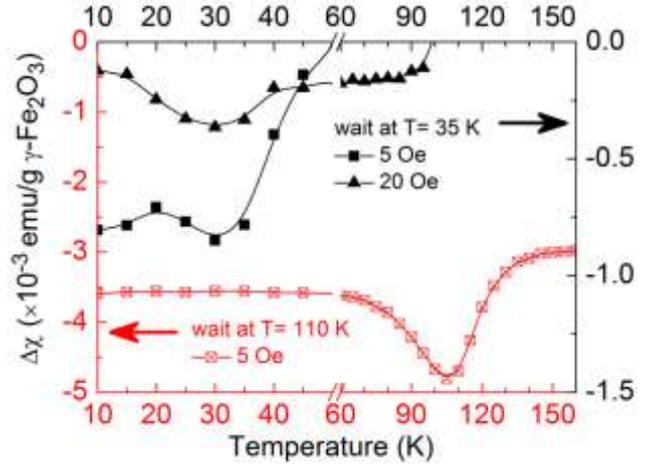

**Fig. 6** Temperature dependence of the difference, $\Delta\chi = \chi_{wait}(T) - \chi_{ref}(T)$, of the ZFC dc susceptibility curves (H= 5 Oe) before and after waiting for $10^4$ sec at 35 K (filled squares) and at 110 K (crossed squares) for a 50.2 nm CNCs powder sample. The memory effect at 35 K was also verified under a field of H= 20 Oe (filled triangles).

Furthermore, the strength of the dipolar interactions has been studied by the Mössbauer spectroscopy. The MS at 200 K (a temperature below the freezing point of CNCs liquid dispersions, as indicated by the SQUID measurements; Fig. S5) for the frozen solutions of the small and large CNCs, is compared with the MS in their powder form where the inter-cluster dipolar interactions might be stronger. Indeed, the MS show qualitatively similar features for the two materials' forms (Fig. S6). With the inter-cluster dipolar interactions being weaker in the frozen solutions, we presume that the superspin glass freezing is associated with the assembly and puckering of the superparamagnetic $\gamma$-Fe$_2$O$_3$ NPs within the CNCs.

At lower temperature ($T_f$), the maximum of $\chi''(T)$, can be attributed either to a surface spin-glass freezing[67] or to a blocking temperature because of a second particle size distribution for example. In order to clarify its origin the likely existence of memory effects was studied again. The memory curve for the small CNCs was measured (H= 5 and 20 Oe), but in this case after having kept the sample at 35 K for $10^4$ sec. The depth of the low-temperature aging curve dip is much weaker, $\Delta\chi\sim 1.0\times 10^{-4}$ emu/g $\gamma$-Fe$_2$O$_3$ (Fig. 6), than that observed at $T_B$. Such a minimal decrease of the susceptibility,

Δχ, is reminiscent of the effect of the surface spin disorder in various nanoparticle systems, including for example, 6 nm $\gamma$-$Fe_2O_3$ NPs[34], 8 nm Ni ferrite[26] and 3 nm Co ferrite[39]. Its possible existence in the CNCs was investigated further through the impact of a dc magnetic field, H, on the system's dynamical behaviour (Fig. 7a). The spin-freezing temperature, $T_f$, was found to decrease with raising magnetic field strength according to Almeida-Thouless law (Fig. 7c).[68] The latter provides good evidence that the spin-freezing at $T_f$ is associated with part of the moments located at the surface of the individual nanoscale units composing each system.[61] Interestingly, as the particle volume fraction (φ) increases, a large shift ΔT (~ $T_f^{cluster} - T_f^{individual}$) of the χ"(T) peak maximum is observed (e.g. ΔT~ 9 K, between φ 0.47 and 0.60, at f= 100 Hz) (Fig. 5a-b) at zero dc field. This behavior is similar to that observed in 6 nm NiO NPs, where the $T_f$ shifts to higher values because of the influence of the increasing strength of the dipolar interactions.[69]

The above experimental evidence together with the literature claims[26, 63] for disordered surface spins justify the description of the frequency dispersion of the χ"(T) maximum by the power-law. In this case the extracted zv values (Table 1; Fig. 5d, f, h) are found comparable to those reported before for surface spin-glass freezing (4< zv< 5), while the attempt time, $\tau_0$ (~ $10^{-4}$ s), is also within the expected range. Worth to be noted that so long $\tau_0$s (~$10^{-5}$-$10^{-6}$ s) have been reported for the surface spin-glass freezing in Ni ferrite[26] and $\gamma$-$Fe_2O_3$ NPs[34]. Therefore on the basis of the previous considerations the contribution of the disordered surface spins in the spin-glass state is not excluded. Such an emerging microscopic physical picture is the subject of evaluation by means of elaborate Monte Carlo simulations in the following section.

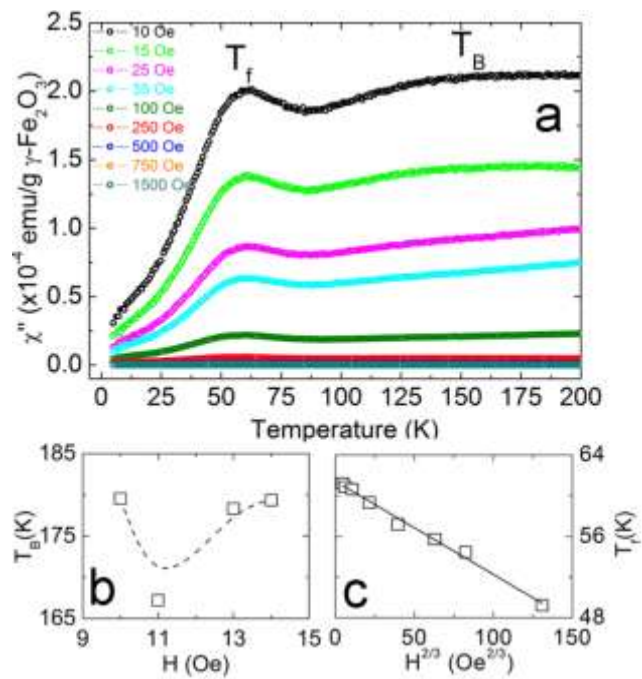

**Fig. 7** (a) The temperature dependence of the imaginary part, χ"(T), of the ac susceptibility under DC fields, ranging from 10 to 1500 Oe, for a 50.2 nm CNCs powder sample. The field dependence of the $T_B$ (dashed line is a guide to the eye) (b) and $T_f$ (line is a fit with Almeida-Thouless law) (c).

**Table 1.** Analysis of the relaxation times, for all three samples (N), on the basis of the Arrhenius, Vogel-Fulcher and power-law phenomenological description (equations 1-3) of the spin dynamics.

| N (nm) | φ | Arrhenius Law | | Vogel-Fulcher Law | | | Power-Law | | |
|---|---|---|---|---|---|---|---|---|---|
| | | $\tau_0$ (s) | $E_a/k_B$ (K) | $\tau_0$ (s) | $E_a/k_B$ (K) | $T_0$ (K) | $\tau_0$ (s) | zv | $T^*$ (K) |
| **Low-temperature maximum** | | | | | | | | | |
| 12.7 | 0.47 | $9.9 \times 10^{-9}$ | 572.8 ± 63.1 | $3.3 \times 10^{-4}$ | 27.5 ± 5.5 | 28.3±0.5 | $9.5 \times 10^{-4}$ | 5.1 ± 0.5 | 25.0 ± 9.3 |
| 50.2 | 0.60 | $4.0 \times 10^{-9}$ | 757.9 ± 63.0 | $6.6 \times 10^{-5}$ | 73.3 ± 8.1 | 32.6±0.4 | $2.4 \times 10^{-4}$ | 4.0 ± 0.2 | 35.5 ± 1.9 |
| 85.6 | 0.72 | $3.8 \times 10^{-8}$ | 627.8 ± 35.7 | $8.6 \times 10^{-6}$ | 186.2 ± 7.5 | 21.6±0.6 | $8.6 \times 10^{-4}$ | 4.3 ± 0.1 | 31.3±0.6 |
| **High-temperature maximum** | | | | | | | | | |
| 12.7 | 0.47 | $5.6 \times 10^{-19}$ | 6623.3 ± 394.5 | $9.2 \times 10^{-7}$ | 405.0 ± 32.4 | 129.7±1.0 | $3.9 \times 10^{-7}$ | 8.8 ± 0.4 | 133.3±16.0 |
| 50.2 | 0.60 | $2.3 \times 10^{-66}$ | 27515.3 ± 3096.7 | $2.1 \times 10^{-7}$ | 137.7 ± 4.0 | 173.9±4.4 | $1.6 \times 10^{-11}$ | 6.5 ± 0.1 | 178.8± 0.8 |
| 85.6 | 0.72 | $3.3 \times 10^{-23}$ | 5237.3 ± 230.2 | $2.3 \times 10^{-9}$ | 546.6 ± 14.1. | 74.0±0.3 | $2.4 \times 10^{-8}$ | 10.0±0.14 | 86.5±2.0 |

## 3.5 Rationalization of Spin-glass Behavior by Monte Carlo Simulations

The two different spin-dynamical regimes identified by the corresponding maxima in $\chi''(T)$ (characterized by differing attempt times, $\tau_0$, on the basis of the power law scaling analysis – Table 1) warrant further insight. In order to examine the mechanism behind the spin-glass behavior, in both individual NPs and CNCs samples, the Monte Carlo method provided simulations of the isothermal (at T= 0.05 in reduced units - see

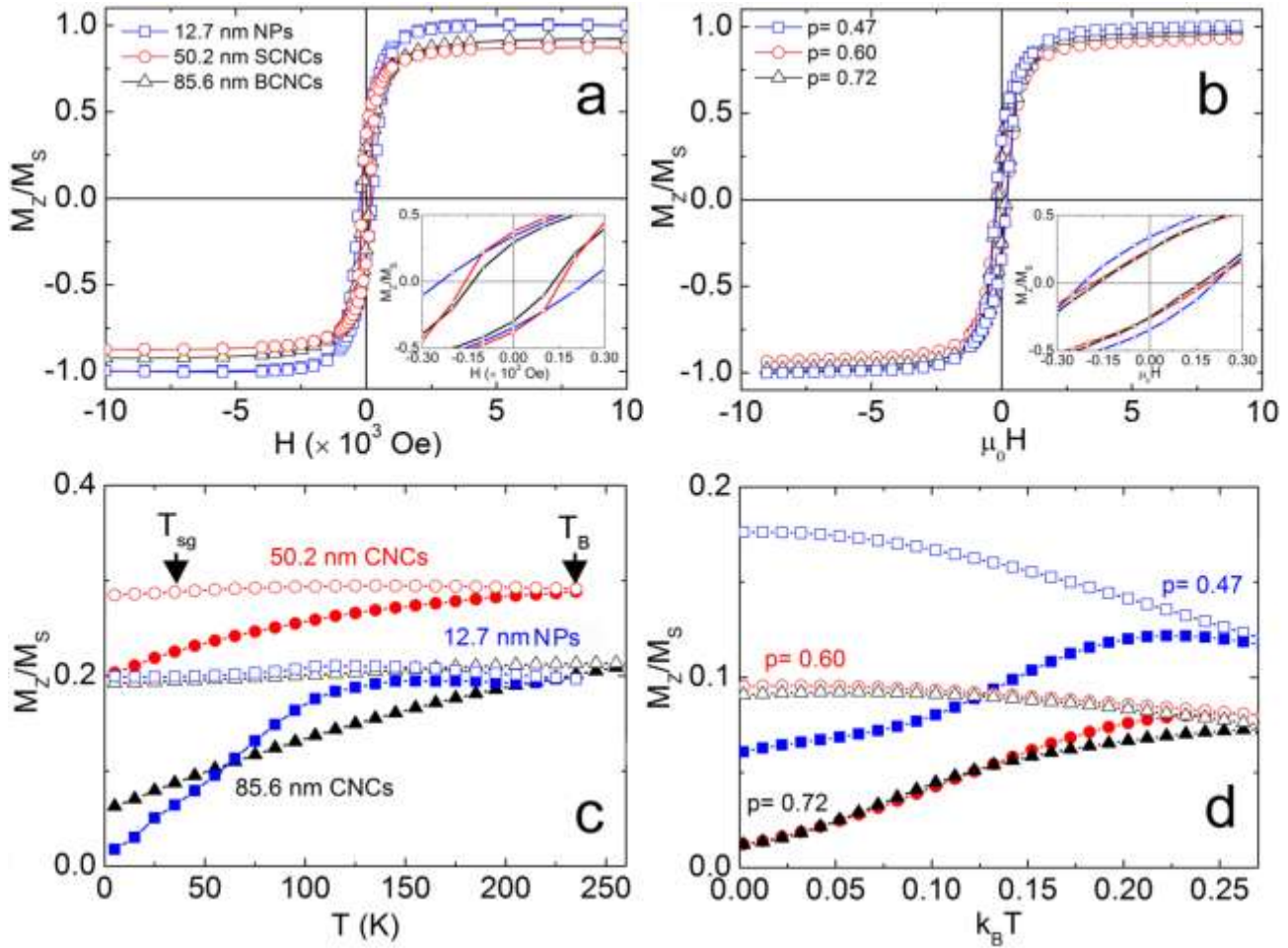

**Fig. 8** Experimental (T= 5 K) (a) and Monte Carlo simulated (T= 0.05) (b) isothermal magnetization curves, for the 12.7 nm individual NPs (squares), 50.2 (circles), the 85.6 nm (triangles) CNCs under zero-field cooling conditions. ZFC (filled points) and FC (open points) measured (c) and simulated (d) magnetization curves as function of temperature for the same samples under a magnetic field of 50 Oe. All the curves are normalized with the saturation magnetization ($M_S$) of the individual NPs; that is 78.6 emu/g $\gamma$-$Fe_2O_3$ for the experimental and 503.2 $\mu_B$ for the simulated results, respectively.

§2.4) magnetization curves under a ZFC procedure and the temperature dependence of the magnetization, M(T), under ZFC and FC protocols. Low-temperature is important here because the contribution of the thermal agitation to the magnetization can be reduced. In general, the morphology, the size (d) of the component NPs and the thickness of the surfactant layer contribute in the determination of the parameters utilized in MC simulations.

In the first model, for the individual NPs with core/surface like morphology and concentration p= 0.47, a common anisotropy easy-axis has been considered for the three spins inside each NP. However, this axis is assumed to be randomly oriented from particle to particle. In addition, the calculations took into account a larger surface anisotropy ($K_S$) and a bigger magnetic moment (m) per particle than in the models that simulated the CNCs (Table 2). These parameters were rationally chosen upon consideration of (a) the different morphology of the individual NPs (not completely spherical by TEM – Fig. 1 – as well as with a broader size distribution, as indicated by TEM and Mössbauer experiments) and (b) the approximate thickness of the surfactant layer (as evaluated by TGA, where the lower weight fraction reflects a thicker organic capping layer; see §2.3.e)[70], which appears to influence the degree of the defected surface coordination environment. The utilized approximation, concerned with the two-domain spin arrangement of the individual NPs, comes along with the understanding that the thicker the surface coordinating organic layer, the lower the disorder of the uncompensated surface spins.[28,71,71] Effectively, this consideration is in agreement with an expected higher saturation magnetization, $M_S$, as indeed observed for the individual NPs (against the CNCs) in their experimental magnetization curves (Fig. 8a).

At the same time in the model which describes the behavior of the CNCs (p= 0.62, 0.70) the core spin anisotropy axis for each nanoparticle contained in a cluster was assumed parallel from site to site, i, effectively resembling the crystallographic alignment of adjacent NPs in a cluster (Fig. 3e, f). On the other hand the surface spin anisotropy at each site ($\hat{e}_i$) was modeled at a random direction with respect to the core (Scheme 1). The disorder between the two domains and the imposed intra-particle interactions ($J_{13}$, $J_{12}$, $J_{23}$) appear as necessary parameters for the MC simulation to fulfill the experimental observation of a reduced saturation magnetization, $M_S$, in the CNCs as compared to the individual NPs (Fig. 8a, b). This behavior is quite the opposite to that reported for multi-core $\gamma$-$Fe_2O_3$ particles (D< 30 nm). In the latter the component units were also crystallographically oriented, but as they were lacking any disorder between the surface spins and the spin of their core no exchange coupling anisotropy was established.[8] Conversely, the CNCs display minimal, but resolvable exchange-bias ($H_{eb}$~ 10 Oe), which is progressively reduced at increasing φ, in analogy to $MnFe_2O_4$@$\gamma$-$Fe_2O_3$ NPs[72] (where $H_{eb}$ goes from 50 Oe at φ~ 0.04 to 25 Oe, at φ~ 0.14), therefore corroborating their subtle internal structural characteristics.

In the above 3-spin models, the dipolar energy strengths, g, were calculated (see §2.4) for the three cases of particle concentration, p (or φ) (Table 2). They were found to follow the trend of the variation of the inter-particle interaction energy, $T_0$, obtained by the Vogel-Fulcher law (Table 1), necessitating the presence of dipolar interactions in all samples.

Overall, the simulations of M(H) as a function of p showed a progressive decrease of $H_C$ and a lowering of $M_S$ (Fig. 8b) in good agreement with the experimental curves (Fig. 8a). Further on, the blocking temperature ($T_B$) was calculated to grow with p and the T-dependent FC magnetization curves become flat (Fig. 8d) at $k_BT \leq 0.05$, similarly to the measured ones (Fig. 8c), indicating the presence of spin-glass dynamics.[3]

**Table 2:** Main parameters used in MC simulations: p (or φ) the concentration of the magnetic material introduced in the model in order to simulate the individual NPs and the CNCs, g the dipolar energy strength, m the mean particle magnetic moment, $K_C$ and $K_{srf}$ the anisotropy energy constants for the core and the surface shell. These parameters are dimensionless as they are normalized by 10 $K_C$.

| N | p | g | m | $K_C$ | $K_{srf}$ |
|---|------|-------|------|-----|------|
| 1 | 0.47 | 0.884 | 1.17 | 0.1 | 3.5 |
| 2 | 0.60 | 0.955 | 0.87 | 0.1 | 2.5 |
| 3 | 0.72 | 0.865 | 0.77 | 0.1 | 2.5 |

With the purpose to verify the accuracy of the first model (p= 0.47), a simulation was carried with only one spin in the core (Stoner-Wohlfarth model) (Fig. S7). The $H_C$ in this case is smaller than in the 3-spin model and the blocking temperature is higher, approaching that of the CNCs. In addition, a simulation assuming a disordered spin arrangement similar to that employed for the CNCs was tested. While the $H_C$ does not change considerably, the blocking temperature increases, approximating that of the CNCs (Fig. S8) in disagreement with the experimental behaviour. Therefore these simulations support the scenario that no considerable spin disorder (due to defected surface coordination environment) would be required for the observed magnetic behavior of the individual NPs.

Furthermore, to identify the factors dictating the spin-glass behavior we have examined the cases, where, either the dipolar interactions (g= 0) (Fig. S9) or the intra-particle spin-exchange interactions ($J_{13}$= $J_{12}$= $J_{23}$= 0) (Fig. S10) were selectively "switched off". When g= 0, the spins inside each nanoparticle tend to spontaneously couple ferrimagnetically (but with randomly oriented easy-axis amongst sites), with effective diminution of their surface spin disorder and as such, the FC M(T) curves present a short plateau (Fig. S9b, d, f - curves with symbols). This is in contrast to the experimentally observed long plateaus in FC- M(T) (Fig. 8c) and the shift of the blocking temperatures to higher values. On the other hand for suppressed intra-particle interactions ($J_{13}$= $J_{12}$= $J_{23}$= 0), the spins inside the nanoparticles do not interact with each other, permitting the surface moments to become decoupled from nearby spins (surface and core) and adopt a random configuration. The randomness in the spin arrangement leads to an increase in the $M_S$ (Fig. S10a, c, e), being highest for the p= 0.72 and a similar, but reduced $T_B$ for all samples (Fig. S10b, d, f - curves with symbols). This second case appears also in contrast to the experimental data corroborating the intra-particle exchange interactions are non-negligible. The previous two tests validate the accuracy of the originally chosen MC models.

Overall, MC simulations suggest that the spin-glass dynamics in the individual NPs are driven mainly by strong dipolar inter-particle interactions, while in their assembled analogues, in the form of a cluster, additional spin disorder due to defected surface coordination environment of the composing NPs is also essential. The extra spin disorder at the surface of the nanoparticles inside each cluster that is not apparent in FFT patterns (Fig. 3e, f), enhances the spin frustration and results in a reduced saturation magnetization compared to that of the individual NPs. For that reason the crystallographic orientation (Fig. 3e, f) that may prejudge for the opposite effect, is not an adequate condition for enhanced magnetization. Knowledge is required on the contribution of the various interaction mechanisms that are operative at much smaller length scales.

## 4. Conclusions

We have presented a modified high-temperature polyol process for the growth of tunable dimension (50 and 86 nm in diameter) colloidal nanoclusters of maghemite. These exhibit good stability and high dispersibility in aqueous media, without any further functionalization. The clusters are composed of crystallographically oriented maghemite nanocrystals of about 13 nm (d) in diameter. As the interest is driven by the evolution of the properties upon the assembly of NPs, individual nanocrystals of comparable size are also prepared under similar conditions.

The impact of the nanoparticle assembly process on the properties, as dictated by the increasing volume fraction, φ (0.47 for individual NPs and 0.60, 0.72 for small and large clusters, respectively), of the inorganic magnetic phase has been investigated. In addition to the room-temperature ferrimagnetism, the extended study of the magnetic dynamics recognizes for all φs two maxima in the temperature dependence of the dissipative part of the ac susceptibility, $\chi''(T)$. They are indentifying features of different spin-dynamical regimes characterized by varying relaxation times. Scaling-law analysis and Monte Carlo simulations suggest that a spin-glass state arises (i) in the individual NPs from strong dipolar interactions and their impact on the surface spin disordering, whereas (ii) in the assembly of such NPs in clusters, with increased φ, from the interplay of dipolar interactions with an additional spin disorder due to the defected nanoparticle surface coordination environment. In this second case, the former interaction mechanism is responsible for the observed memory effects at the high-temperature superspin glass transition ($T_B$). Whilst, the additional surface spin disorder generates a much weaker, low-temperature ($T_f$) memory effect and a resolvable exchange-bias.

This study illustrates how the knowledge and contribution of the different length-scale microscopic mechanisms are crucial for the development of technologically useful nanoparticle assemblies. Designing such materials with optimum characteristics may address particular magnetically driven applications, such as, in diagnosis (MRI), therapy (hyperthermia treatment) or as data storage (memory technologies).

**Acknowledgements**
This work was supported by the European Commission through the Marie-Curie Transfer of Knowledge program NANOTAIL (Grant no. MTKD-CT-2006-042459). L.M. acknowledges financial support by the Italian FIRB grant (contract #RBAP115AYN). M.V and K.N.T acknowledge financial support from the European Social Fund (EU) and Greek national funds through the Operational Program "Education and Lifelong Learning" in the framework of ARISTEIA I (Project COMANA/ 22). A.L. thanks the Italian FIRB RINAME and Fondazione Cariplo project (no. 2010-0612).
73

**Notes and references**

[a] Institute of Electronic Structure and Laser, Foundation for Research and Technology-Hellas, Vassilika Vouton, Heraklion 71110, Greece.

Fax: +30 2810 301305; Tel: +30 2810 391344; E-mail: lappas@iesl.forth.gr

[b] Department of Physics, Aristotle University of Thessaloniki, 54124 Thessaloniki, Greece.

[c] IAMPPNM, Department of Materials Science, NCSR "Demokritos", Aghia Paraskevi, 15310, Athens, Greece.

[d] Department of Physics, University of Ioannina, Ioannina 45110, Greece.

[e] Dipartimento di Fisica, Università degli studi di Milano and INSTM, Via Celoria 16, I-20133 Milano, Italy.

[f] Istituto Italiano di Tecnologia, Via Morego 30, 16163 Genova, Italy.

† Electronic Supplementary Information (ESI) available: DLS and z-potential curves for CNCs; Table of measured particle diameters by DLS and TEM; Mössbauer spectra for CNCs at different temperatures; Table of the Mössbauer spectra fitted parameters; analysis of the frequency dispersion of the $\chi''(T)$ maximum by the Arrhenius, and Vogel-Fulcher laws; ZFC-FC dc susceptibility curves of the CNCs in solutions; Mössbauer spectra at 200 K in solution and powder form; Monte Carlo simulations of the magnetization for individual nanoparticles assuming a Stoner-Wohlfarth model; Monte Carlo simulations of the magnetization for individual nanoparticles assuming the CNCs model; Monte Carlo simulations with g= 0 and $J_{12}=J_{13}=J_{23}=0$.